\documentclass[11pt]{article} 

\usepackage[utf8]{inputenc} 
\usepackage{caption}

\usepackage{geometry} 
\usepackage{tikz}

\usepackage{graphicx} 

\usepackage{booktabs} 
\usepackage{array} 
\usepackage{paralist} 
\usepackage{verbatim} 
\usepackage{subfig} 

\usepackage{fancyhdr} 
\usepackage{sectsty}
\allsectionsfont{\sffamily\mdseries\upshape} 

\usepackage[nottoc,notlof,notlot]{tocbibind} 
\usepackage[titles,subfigure]{tocloft} 

\usepackage{graphicx}
\usepackage[intlimits]{amsmath}
\usepackage{latexsym}
\usepackage{amsfonts}
\usepackage{amssymb}%
\makeatletter
\newfont{\footsc}{cmcsc10 at 8truept}
\newfont{\footbf}{cmbx10 at 8truept}
\newfont{\footrm}{cmr10 at 10truept}
\newtheorem{theorem}{Theorem}

\newtheorem{corollary}[theorem]{Corollary}

\newtheorem{definition}[theorem]{Definition}

\newtheorem{lemma}[theorem]{Lemma}

\newtheorem{proposition}[theorem]{Proposition}
\newtheorem{remark}[theorem]{Remark}

\newcommand{\proofstart}{{\bf Proof\hspace{2em}}}

\newcommand{\proofend}{\hspace*{\fill}\mbox{$\Box$}}

\title{Wheel Random Apollonian Graphs}
\author{
        Piero Giacomelli\footnote{pgiacome@gmail.com} \\
                Dipartimento di matematica\\
        Universit\`a di Padova
}

\date{} 

\begin{document}
\maketitle
\begin{abstract}
In this paper a subset of  High-Dimensional Random Apollonian networks, that we called Wheel Random Apollonian Graphs (WRAG), is considered. We  show how to generate a Wheel Random Apollonian Graph from a wheel graph. We analyse some basic graph properties like vertices and edges cardinality, some question concerning cycles and the chromaticity in such type of graphs, we suggest further work on this type of graphs.
\end{abstract}
\begin{flushleft}
To Michela.
\end{flushleft}
\section{Introduction}
\label{Introduction}
Random Apollonian Networks \cite{2004cond.mat..9414Z} (RANs), have been introduced to show how a simple generating rule can create networks with a very large clustering coefficient and very small average distance. Starting from a triangle graph at step $0$ at step $1$ , a new vertex is added inside the triangle and is connected with  three new edges to the three vertex .
The result is a graph that has three triangles as subgraphs, we choose randomly one of these not chosen before and repeat the step $0$. This simple iteration process generate a Random Apollonian Network (RAN). There have been other works that studied the distribution of distances\cite{2007arXiv0712.2129B} and  the degree correlations  of these graphs\cite{2007PhyA..380..621Z}. In Zhang\cite{2006PhyA..364..610Z} a generalization of the  growing process has been established by introducing the concept of High-dimensional random Apollonian networks (HDRANs). Generalizing the procedure, for an High-dimensional random Apollonian network the starting point is a complete graph $k_{d+2}$\footnote{$k_{d+2}$ is defined as a graph without
loops whose $d+2$ vertices are pairwise adjacent see \cite{West.IntroductiontoGraph}}. At every new step a new vertex is added to a randomly chosen subgraph isomorphic to a $d+1$-click graph, not chosen before. Finally the new vertex is connected to all the vertices of the chosen subgraph. So (RANs) are a special case of HDRANs, in the same paper\cite{2006PhyA..364..610Z} the degree distribution and the clustering coefficient of HDRANs have been detailed.
The aim of those notes is to focus on a particular case of HDRANs, the ones that have a wheel graph as starting seed, we called them Wheel Random Apollonian Graphs (WRAGs). These graphs are a particular case of HDRANs, even from the starting graph even for the subgraph added at iteration step which is, following the growing process, a particular wheel graph. 
This paper is organized as follow: on section one we recall some  preliminaries on wheel graph and we detail the construction of Wheel Random Apollonian graphs.  Section two is dedicated to show some results on WRAG concerning cardinality of vertices and edges and some notes on cycles. Section three will deal with chromaticity. Last section describes some open problems and questions that arise from Wheel Random Apollonian Graphs.

\section{Basic Definitions}
\label{Basic Definitions}

Let us start with some basic definitions 
\begin{definition}
A wheel graph $W_m(v_0), m \geq 4, m \in \mathbb{N}$  with hub in $v_0$ of order $m$  is an planar, undirected, complete graph $G=(V,E)$ where 
\begin{equation*}
V=\{v_0,v_1,v_2,\dots , v_{m-1}\}
\end{equation*} is the set of vertex.
\begin{equation*}
E=\{(v_i,v_j) \subseteq V\times V\}
\end{equation*} is the set of edges such that
\begin{equation*} 
E = S \cup C
\end{equation*} where 
\begin{equation*} 
S = \{(v_0,v_i) \subset V\times V| i=1\dots m-1\}
\end{equation*} 

 and
\begin{equation*} 
C =\{(v_i,v_{i+1}) \subset V \times V | i=1\dots m-2\}\cup(v_{m-1},v_1).
\end{equation*} 
  
\end{definition}
Alternatively we can define $W_m(v_0)$ as a planar undirect graph with $m$ vertices and no loops formed by connecting the vertex $v_0$ (the hub) to all vertices of a $(m-1)$-cycle graph. By definition, $W_m(v_0)$ is connected and complete  $|V| = m$ and $|E| = |S \cup C| = |S| + |C|$ ($S\cap C = \emptyset$) and so $|E| = (m-1) + (m-1) = 2(m-1)$. It is also easy to see that for $m \in \mathbb{N}, m \geq 4$ giving any two vertex $v_i,v_j$ the number of edges connecting $v_i,v_j$, i.e. is a function:
\begin{equation*} 
V\times V\longrightarrow \mathbb{N}
\end{equation*} 
\begin{equation*} 
P_{ij}(v_i,v_j) = 
\begin{cases}
1 \text{ if } (v_i,v_j) \in $C$ \\
2 \text{ if } (v_i,v_j) \notin $C$
\end{cases}
\end{equation*} 
because by definition the path form $(v_i,e_{i0},v_0,e_{0j},v_j)$ is always an admittable path. So the diameter of $W_m(v_0)$, defined as $D=max(P_{ij}) = 2$.
Before proceeding with the definition of Wheel random apollonian graph, we recall that:
\begin{definition}
Let $G$ a graph an $n$-cycle is a closed, simple, path, with $n-1$ distinct vertices and one vertex repeated that is the starting and the ending vertex of the $n$-cycle.
\end{definition}
For the iteration step of a wheel random apollonian graph is is important to define:
\begin{definition}
let $G$ be a graph we define $T(G)$ as the set of all $3$-cycle of $G$. 
\end{definition}
Clearly if  $G = W_m(v_0)$
\begin{equation*} 
T(G) = T(W_m(v_0)) = \{\Delta_{i,0,k} | \forall i=1\dots m-1, k = 2\dots m-2 \} \cup \Delta_{m-1,0,1}
\end{equation*} 
where 
$\Delta_{i0k}$ is the $3$-cycle that touch the vertex $v_i,v_0,v_k$. By construction $|T(W_m(v_0))| = m-2+1 = m-1$.
\newline 
Let now introduce a procedure for building an high dimension random apollonian graph starting from a wheel graph. This procedure follows the one in the paper by Zhang\cite{2006PhyA..364..610Z} in the iteration step, however it differs from the starting seed and the type of graph added at every iteration step. First we  recall \cite{2006PhyA..364..610Z}
\begin{definition}
Let $K_{d+2}$ a complete graph\cite{West.IntroductiontoGraph}. An High Dimension Random Apollonian Graph (HDRAG) $A(d,t)$ where $d,t \in \mathbb{N},d > 0, t > 0$ is a recurrent sequence of graph definened as follows
\begin{equation*} 
A(d,t) = 
\begin{cases}
A(d,0) = K_{d+2} \\
A(d,t + 1) = A(d,t) \cup K  
\end{cases}
\end{equation*} 
where $K$ is the graph obtained by choosing a subgraph $S$ of  $A(d,t)$, isomorphic to a $d+1$-clique, not selected before, adding a new vertex on it  and connecting it to other vertices of $S$. 
\end{definition}
So using this, we can define the following:
\begin{definition}
let $W_m(v_0)$ be a wheel of order $m$ with hub in $v_0$, a Wheel Random Apollonian Graph Sequence $WRAGS_i(0,m)$, $i, m \in \mathbb{N}, m\geq 4,i > m$, is a recurrent sequence of graphs defined as follows
\begin{multline} 
WRAGS_i(W_m(v_0)) = 
\begin{cases}
WRAGS_0(W_m(v_0)) = W_m(v_0) \\
WRAGS_{i+1}(W_m(v_0)) = WRAGS_{i}(W_m(v_0))  \cup W_4(v_{i+1})
\end{cases}
\end{multline} 
where $v_{i}$ is a new vertex that became the hub of a $W_4(v_{i+1})$ wheel, where the other vertices are the ones of a random choosen $3$-cycle of $T(WRAGS_{i}(W_m(v_0)))$, not chosen in the previous steps.
\end{definition}
So starting by a wheel $W_m(v_0)$ we choose randomly a $3$-cycle of $T(W_m(v_0))$ not chosen  before, and we add a new vertex inside it and whe use this as a center for a new $W_4(v_{i+1})$ wheel, to obtain a new graph.  

We are now ready to give the main
\begin{definition}[Wheel Random Apollonian Graph]
A Wheel Random Apollonian Graph $WRAG(m,n)$ of seed $m$ and order $n$, $m,n \in \mathbb{N}, 4\leq m < n$ is Wheel Random Apollonian Graph Sequence ($WRAGS$) stopped at the $i=n-m$ step,formally:
\begin{equation*} 
WRAG(m,n) =  WRAGS_{n-m}(W_m(v_0)).
\end{equation*} 
\end{definition}
So by definition a random apollonian network is a wheel random apollonian graph with $m=4$. Always by definition Wheel Random Apollonian Graph Sequence is an of (HDRAN) by choosing $d+2=m$ and every new step choosing a $3$-cycles graph.
\newline
To see if the definition is well posed, we need the following
\begin{lemma}
\label{NumberOf3cycle}
If $m,i \in \mathbb{N}$ such that $ m \geq 4, i \geq 0  $  then 
\begin{equation}
|T(WRAGS_{i}(W_m(v_0)))| = 
\begin{cases}
	m + 3i \text{ if } m = 4 \\
	m-1 + 3i \text{ if } m > 4 
\end{cases}
\end{equation}. 
\end{lemma}
\proofstart
By induction on $i$. 
\begin{itemize}
	\item {$i=0$. In this case if $m=4$ then $T(WRAGS_i(W_m(v_0))) = T(W_4(v_0)) = m$ because we count $m-1$, $3$-cycles having vertex in $v_0$ and a $3$-cycle that pass through the vertices $v_1,v_2,v_3$.   Otherwise if $m>4$ then $T(WRAGS_{m,i}) = T(W_m(v_0)) = m-1$, because we count  $m-1$, $3$-cycles.} 
	\item {$i>0$. The induction hypothesis is $T(WRAGS_{i}(W_i(v_0))) = m-1 + 2i$. Let us analyse  $T(WRAGS_{i+1}(W_m(v_0)))$. Considering the $3$-cycle of vertices $v_r,v_s,v_t$ chosen for passing from the step $i$ to the step $i+1$. It is easy to see that 
\begin{align*}
T(WRAGS_{i+1}(W_m(v_0))) &= \\
												&= T(WRAGS_{i}(W_m(v_0))) \cup \Delta_{v_r,v_{i+1}, v_s} \cup \Delta_{v_s,v_{i+1}, v_t} \cup \Delta_{v_t,v_{i+1}, v_r}
\end{align*}
	where $\Delta_{v_r,v_{i+1}, v_s}$ is the $3$-cycle touching the vertices $v_r,v_{i+1},v_s$, $\Delta_{v_s,v_{i+1}, v_t}$ is the $3$-cycle touching the vertices $v_s,v_{i+1},v_t$ and finally $\Delta_{v_t,v_{i+1}, v_r}$ is the $3$-cycle with vertices $v_t,v_{i+1},v_r$.	
	
So now whe have two cases
\begin{itemize}
	\item{$m=4$. in this case 
		\begin{align*}
		|T|   &= |T(WRAGS_{i+1}(W_m(v_0)))| \\
		      &= |T(WRAGS_{i}(W_m(v_0))) \cup \Delta_{v_r,v_{i+1}, v_s}  \cup \Delta_{v_r,v_{i+1}, v_s} \cup \Delta_{v_t,v_{i+1}, v_r} | \\
		      &= |T(WRAGS_{i}(W_m(v_0)))|  + | \Delta_{v_r,v_{i+1}, v_s}  |  + | \Delta_{v_r,v_{i+1}, v_s} |  + | \Delta_{v_t,v_{i+1}, v_r} |  \\      
		      &= m + 3i  + 1  + 1 + 1  \\
		      &= m + i+  3 \\
		      &= m + 3(i+1).      
		\end{align*}
	}
	\item{$m>4$. The case is similar to the previous one except for that we have a $3$-cycle less so 
		\begin{align*}
			|T| &= |T(WRAGS_{i+1}(W_m(v_0)))| \\
		      &= | T(WRAGS_{i}(W_m(v_0))) \cup \Delta_{v_r,v_{i+1}, v_s}  \cup \Delta_{v_r,v_{i+1}, v_s} \cup \Delta_{v_t,v_{i+1}, v_r} | \\
		      &= | T(WRAGS_{i}(W_m(v_0))) |  + | \Delta_{v_r,v_{i+1}, v_s}  |  + | \Delta_{v_r,v_{i+1}, v_s} |  + | \Delta_{v_t,v_{i+1}, v_r} |  \\      
		      &= m -1+ 3i  + 1  + 1 + 1  \\
		      &= m-1 + i+  3 \\
		      &= m-1 + 3(i+1).      
		\end{align*}		
	
	}
\end{itemize}
in both case the sum is direct because $\Delta_{v_r,v_{i+1}, v_s}   \cap T(WRAGS_{i}(W_m(v_0))) = \emptyset,  \Delta_{v_r,v_{i+1}, v_s} \cap T(WRAGS_{i}(W_m(v_0))) = \emptyset,   \Delta_{v_t,v_{i+1}, v_r} \cap T(WRAGS_{i}(W_m(v_0))) = \emptyset$. 

This complete the proof.
	} 
\end{itemize}
\proofend
\newline
Using this lemma, it is possible to give the following 
\begin{proposition}
If $m,n \in \mathbb{N}$ such that $ m \geq 4 $ and $m<n$ the definition of $WRAG(m,n)$ is well posed. 
\end{proposition}
\proofstart
By \ref{NumberOf3cycle} at every step $i$ of a $WRAGS_{i}(W_m(v_0))$ the set $T(WRAGS_{i}(W_m(v_0))) \neq \emptyset$, and there are $3$ new chooseable $3$-cycle and from the previous total count of $3$-cycle we have to leave the one already chosen in the iterative step, so there is at least two chooseable $3$-cycle not chosen before.
\proofend
\newline
As a consequence of \ref{NumberOf3cycle} we have that
\begin{corollary}
$T(WRAG(m,n)) = 3n-8$ if $m=4$ or $T(WRAG(m,n)) = 3n-2m-1$ if $n>4$ .
\end{corollary}
\proofstart
\begin{itemize}
	\item{If $m=4$, then $T(WRAG(m,n)) = T( WRAGS_{n-m}(W_m(v_0))) = m + 3(n-m) = 3n-2m = 3n-8.$ }
	\item{If $m>4$, then $T(WRAG(m,n)) = T( WRAGS_{n-m}(W_m(v_0))) = m-1 + 3(n-m) = 3n-2m-1.$ }
\end{itemize}
\proofend
\newline
It is also interesting to notice that by costruction the girth, i.e. the length of the shortest cycles, of a Wheel Random Apollonian Graph is $3$. 
\newline
Reconsidering lemma  \ref{NumberOf3cycle} at any iteration step the number of chooseable $3$-cycles that can be used for the next iteration step increase with $3$ new $3$-cycles but decrease of one $3$-cycle that was chosen in the step before. So at every iteration step the total number of chooseable $3$-cycles increase by two. This the generalization of same fact showed for RANs\cite{2004cond.mat..9414Z}.  If we define the number of chooseable triangles in a Wheel Random Apollonian Graph Sequence at a generic step $i$ as $C(WRAGS_i(W_i(v_0)))$ we have that $C(WRAGS_i(W_m(v_0))) = m-1 +2i$.  \newline
This fact can be used to show that giving $m,n \in \mathbb{N}$ with $m \geq 4 $ and $m < n$, the number of possible different Wheel Random Apollonian Graphs that we can construct giving $m,n$ are 
\begin{equation}
\prod_{i=0}^{n-m} C(WRAGS_i(W_m(v_0))) = \prod_{i=0}^{n-m} m-1 + 2i.
\end{equation}
Using the computational knowledge engine wolframalpha\footnote{http://www.wolframalpha.com} we where able to evaluate the product finding that
\begin{equation}
\prod_{i=0}^{n-m} m-1 + 2i = (m-1)2^{n-m}\biggl( \frac{m+1}{2} \biggl)_{n-m} =	\frac{2^{n-m+1} \Gamma \biggl(n-\frac{m}{2} + \frac{1}{2}  \biggl)}{\Gamma \biggl(\frac{m}{2} - \frac{1}{2}  \biggl)} 
\end{equation}
where $\biggl( \frac{m+1}{2} \biggl)_{n-m}$ is the pochhammer symbol\cite{Abramowitz} and $\Gamma$ is the Gamma function.
So we have
\begin{proposition}
Let $n,m \in \mathbb{N}$ with $m \geq 4, m<n$ then  the possible $WRAG(n,m)$ that can be created are 
\begin{equation}
(m-1)2^{n-m}\biggl( \frac{m+1}{2} \biggl)_{n-m} 
\end{equation}
or alternately
\begin{equation}
	\frac{2^{n-m+1} \Gamma \biggl(n-\frac{m}{2} + \frac{1}{2}  \biggl)}{\Gamma \biggl(\frac{m}{2} - \frac{1}{2}  \biggl)} 
\end{equation}
where $(a)_{j}$ is the pochhammer symbol and $\Gamma(a)$ is the Gamma function.
\end{proposition}

In Figure 1 we can see a possible $WRAG(4,7)$,
\begin{figure}
\centering
\begin{tikzpicture}[scale=3.5]
\tikzstyle{every node}=[draw,shape=circle];
\path (0:0cm) node (v0) {$v_0$};
\path (90:1cm) node (v1) {$v_1$};
\path (90-120:1cm) node (v2) {$v_2$};
\path (90-240:1cm) node (v3) {$v_3$};
\path (90-60:0.4cm) node (v4) {$v_4$};
\path (90-20:0.35cm) node (v5) {$v_5$};
\path (90-180:0.35cm) node (v6) {$v_6$};
\draw 
(v0) -- (v1)
(v1) -- (v2)
(v0) -- (v2)
(v2) -- (v3)
(v0) -- (v3)
(v3) -- (v1)
(v4) -- (v0)
(v4) -- (v1)
(v4) -- (v2)
(v5) -- (v0)
(v5) -- (v4)
(v5) -- (v1)
(v6) -- (v3)
(v6) -- (v0)
(v6) -- (v2)

;
\end{tikzpicture}%
\caption{$WRAG(4,7)$}
\end{figure}
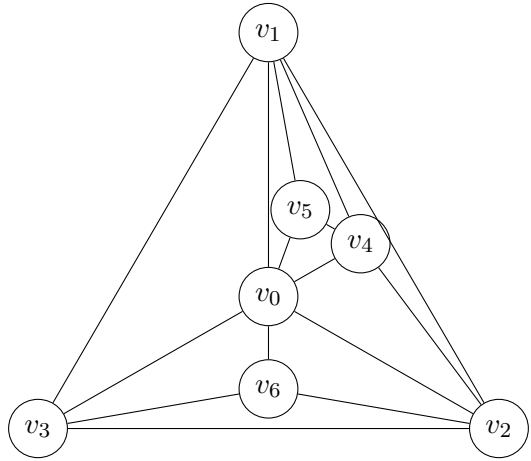
\newline
in Figure 2  we can see a possible $WRAG(5,9)$ .
\begin{figure}
\centering
\begin{tikzpicture}[scale=3.5]
\tikzstyle{every node}=[draw,shape=circle];
\path (0:0cm) node (v0) {$v_0$};
\path (45:1cm) node (v1) {$v_1$};
\path (45-90:1cm) node (v2) {$v_2$};
\path (45-180:1cm) node (v3) {$v_3$};
\path (45-270:1cm) node (v4) {$v_4$};
\path (0:0.5cm) node (v5) {$v_5$};
\path (0-90:0.5cm) node (v6) {$v_6$};
\path (0-180:0.5cm) node (v7) {$v_7$};
\path (0-270:0.5cm) node (v8) {$v_8$};
\draw 
(v0) -- (v1)
(v0) -- (v2)
(v0) -- (v3)
(v0) -- (v4)
(v1) -- (v2)
(v2) -- (v3)
(v3) -- (v4)
(v4) -- (v1)

(v5) -- (v0)
(v5) -- (v1)
(v5) -- (v2)

(v6) -- (v0)
(v6) -- (v3)
(v6) -- (v2)

(v7) -- (v0)
(v7) -- (v3)
(v7) -- (v4)

(v8) -- (v0)
(v8) -- (v1)
(v8) -- (v4)

;
\end{tikzpicture}%
\caption{ $WRAG(5,9)$}
\end{figure}
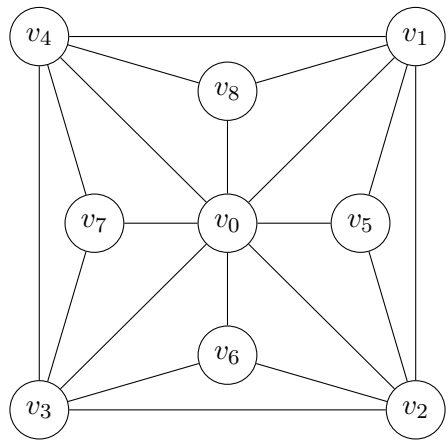
\newpage
%

\section{Basic properties}
\label{Basic properties}
In what follow $WRAG(m,n)$ will be Wheel Random Apollonian Graph where $m,n \in \mathbb{N}$ where $m \geq 4, n>m$. $V$ will indicate the set of vertices, $E$ the set of edges of the $WRAG(m,n)$.

\begin{proposition}
Let $WRAG(m,n)$ be a Wheel Random Apollonian Graph than

\begin{enumerate}
	\item{$|V| = n$.} 
	\item{$|E| = 3n-m-2$.} 
\end{enumerate}
\end{proposition}
\proofstart
\begin{enumerate}
	\item{The starting point of a $WRAG(W_m(v_0)$ is a wheel graph that has $m$ vertices at every step we add to the set of vertex $V$ a new vertex till the $n-m$ final step so we totally have $m+(n-m) = n$ vertices.} 
	\item{ By induction on $i$, we prove that the number of edges of a Wheel Random Apollonian Graph Process $WRAGS_{i}(W_0(m))$  with seed $W_m(v_0)$ is $2(m-1) + 3i.$ 
\begin{itemize}
	\item{$i = 0$. In this case $WRAGS_{i}(W_0(m)) = W_m(v_0)$, so the edges are $2(m-1)$.}
	\item{$i > 0$. In this case $WRAGS_{i+1}(W_0(m)) = WRAGS_{i}(W_0(m)) \cup = W_4(v_{i+1})$. So by induction hypothesis the number of edges is $2(m-1) + 3i + 3 = 2(m-1) +3(i+1)$ }
\end{itemize}
		 } 	
\end{enumerate}
So giving $WRAG(m,n) = WRAGS_{n-m}(W_0(m)) = 2(m-1)+3(n-m) = 2m -2 + 3n - 3m = 3n -m + 2$
\proofend
\newline

A question that arises when referring to cycles in a graph, is if there exist an Hamiltonian cycle. We answer positively to this question in Wheel Random Apollonian Graph with the following 
\begin{theorem} 
\label{WheelAreHamiltonian}
Let $WRAG(m,n)$ a Wheel Random Apollonian Graph of order $n$ with seed $m$, then $WRAG(m,n)$ is Hamiltonian.
\end{theorem} 
We recall that 
\begin{definition} 
Let $G=(V,E)$ a graph. Then a cycle is called Hamiltonian if touch all the vertices of $G$. In this case $G$ is called Hamiltonian.
\end{definition} 
And also we remark that
\begin{remark}
\label{wheel} 
Let $W_m(v_0)$ a wheel graph with hub in $v_0$ then $W_m(v_0)$ is Hamiltonian. 
\end{remark} 
This because the path that touchs the vertices $v_0,v_1,v_2,...v_{m-1},v_0$ is a hamiltonian cycle.
Recalling this facts we are now ready to prove \ref{WheelAreHamiltonian}.
\newline
\proofstart
Again we will show the results as a corollary of this more general results. Giving a sequence of Wheel Random Apollonian Graphs $WRAGS_i(W_m(v_0))$ with seed $W_m(v_0)$ we will show that for each $i \in \mathbb{N}, i \geq 0$ then $WRAGS_i(W_m(v_0))$ contains an Hamiltonian cycle. Using induction
\newline
$i=0$. just proved by \ref{wheel}.
\newline
 $i+1$ let us assume that at step $i$ the $WRAGS_i(W_m(v_0))$ is Hamiltonian, so there is a cycle that touch all the vertex of $WRAGS_i(W_m(v_0))$. At step $i+1$. whe consider the choosen $3$-cycle choosen from $T(WRAGS_i(W_m(v_0)))$. We consider the three vertices of this $3$-cycle let us name them $v_r,v_s,v_t$. By induction al least one on the edges $(v_r,v_s),(v_s,v_t),(v_t,v_r)$ is in the Hamiltonian cycle of the graph $WRAGS_i(W_m(v_0))$, without loss of generality we can choose $(v_r,v_s)$ as this edge. So now starting from the Hamiltonian cycle that touch the vertices $v_0,\dots,v_r,v_s,\dots,v_0$ whe can create a new cycle $v_0,\dots,v_r,v_{n+1},v_s),\dots,v_0$, that by definition touch the new vertex $v_{n+1}$ only once so this cycle is Hamiltonian. So for every step $i$ a   $WRAGS_{i+1}(W_m(v_0))$ is Hamiltonian. So by consequence choosing $i=n-m$ then $WRAG(m,n)$ is Hamiltonian.
\proofend 
\section{Chromaticity of Wheel Random Apollonian Graph}
We remember that the chromatic number of a graph $G=(V,E)$ is the number of different color needed to color each vertex such that giving to connected vertices they have different colors. 
We also remark\cite{} 
\begin{remark}
Let $W_m(v_0)$ a wheel graph with $m \in \mathbb{N}, m \geq 4$ then the chromatic number is 
\begin{equation}
\chi(W_m(v_0)) = 
\begin{cases}
	3 \text{ if } \; m \text{ odd } \\
	4 \text{ if } \; m \text{ even }
\end{cases}
\end{equation}
\end{remark}
It is well known that any planar graph can be colored with at least four colors\cite{2009arXiv0905.3713X}. In case of WRAG we will show that even if the a wheel graph of odd order can be colored with $3$ colors, for a wheel random apollonian graph we always need $4$ colors.
we are now ready to prove the following:
\begin{theorem}
let $WRAGS_i(W_m(v_0))$ a wheel random apollonian graph sequence with $m,i \in \mathbb{N}, m \geq 4, i \geq 0$, then $\chi(WRAGS_i(W_m(v_0))) = 4.$
\end{theorem}
\proofstart
Let consider the two case for the starting $W_m(v_0)$.
\begin{itemize}
	\item{ $m$ odd. in this case at least $\chi(W_m(v_0)) = 3$. Let consider the first step $i=1$, in this case the new vertex $v_{i+m}$ can't be colored with one of the $3$ colors because the $3$-cycle chosen to add this node had three vertices connected with three different colors one on every vertex. So we need to have a fourth color to make every edge with vertices of different colors. If we consider the next step we can choose another $3$-cycle that can have vertices of four different colors. So when we add a the new vertex we are obliged to choose as color the four which is not in the three colors chosen in the $3$-cycle. It is clear so that when we add a new vertex we do not need to add more colors but only to choose the one that is not the $3$-cycle vertices choice.
	   } 
\item{ $m$ even. In this case there is no need to add any color beacause every time we choose a $3$-cycle we have another color that is free for coloring the new vertex added inside that $3$-cycle to create the new $W_4(v_{i+m})$ subgraph.}

\section{Open questions}
\label{Open questions}
In these short notes we addressed some basic properties on wheel random apollonian graphs. As we saw induction is a key factor for finding properties of these type of graphs. There are some interesting questions that arise naturally. We know that giving a generic wheel $W_m(v_0)$ that there's are $m^2 - 3m + 3 $ cycles. If we consider the first induction step and consider the new wheel subgraph $W_4(v_{m+1})$ with hub in $v_{m+1}$, than we add $4^2-3\cdot 4 + 3 -1$ new cycles  but also for every cycle that pass throught the edge that form the cycle of $W_4(v_{m+1})$ subgraph we add $3*2$ more cycles. If would be nice to find a formula for correctly count the total number of cycles.
We study here the chromaticity of $WRAGS_i(W(v_0))$, but again it would be interesting using induction to find the chromatic polynomial of a generic $WRAGS_i(W(v_0))$, knowing that for a generic wheel graph the chromatic polynomial is $P_{W_n}(x)=x((x-2)^{(n-1)}-(-1)^{n}(x-2))$\cite{Godsil} \ if $x$ is the number of possible colours.
Finally it would be interesting to study in a generic Wheel Random Apollonian Graphs the eigenvalues ad eigenvectors of the corresponding adjcency matrix. Observing that at every iteration step of the wheel random apollonian graph sequence we add to the adjacency matrix associated to the previous step a column where every entry is zero apart from three of them which are one. It would be interesting to study how the eigenvalues changes and if it possible to find a relation between adding a new column with 3 entry that are 1 and the eigenvalues of the new matrix.
This topics will be addressed in future works.
\newpage

\bibliographystyle{plain}
\bibliography{gran2} 
	   
\end{itemize}

\end{document}